\documentclass[doublecol]{epl2}

\usepackage{amssymb,amsmath}
\usepackage{graphicx}
\usepackage{dcolumn}
\usepackage{bm}
\usepackage{color}
\usepackage{float}
\usepackage{hyperref}
\usepackage[version=4]{mhchem}
\usepackage{relsize}
\usepackage{comment}
\usepackage{media9}

\title{Reservoir computing based on solitary-like waves dynamics of film flows: a proof of concept}
\shorttitle{Reservoir computing based on solitary-like waves dynamics} 

\author{Ivan S.~Maksymov\inst{1} \and Andrey Pototsky\inst{2}}
\shortauthor{I.~S.~Maksymov and A.~Pototsky}

\institute{
  \inst{1} Artificial Intelligence and Cyber Futures Institute, Charles Sturt University, Bathurst, NSW 2795, Australia\\
  \inst{2} Department of Mathematics, Swinburne University of Technology, Hawthorn, Victoria 3122, Australia
}

\abstract{
Several theoretical works have shown that solitons---waves that self-maintain constant shape and velocity as they propagate---can be used as a physical computational reservoir, a concept where machine learning algorithms designed for digital computers are replaced by analog physical systems that exhibit nonlinear dynamical behaviour. Here we propose and experimentally validate a novel reservoir computing (RC) system that for the first time employs solitary-like (SL) waves propagating on the surface of a liquid film flowing over an inclined surface. We demonstrate the ability of the SL wave RC system (SLRC) to forecast chaotic time series and to successfully pass essential benchmark tests, including a memory capacity test and a Mackey-Glass model test.}

\begin{document}

\maketitle

\section{Introduction\label{sec:1}}
Reservoir computing (RC) \cite{Nak21} and its foundational concepts of context of reverberated input histories \cite{Kir91}, Liquid State Machine (LSM) \cite{Maa02} and Echo State Network (ESN) \cite{Luk09} underpin an important class of machine learning (ML) algorithms that can perform certain functions of a biological brain \cite{Maa02} and that can be trained to recognise patterns and to forecast the response of nonlinear dynamical systems that exhibit chaotic behaviour \cite{Gau21}. This functionality enables using RC systems in many areas of science and technology, including physics, psychology and finance \cite{Mar04, Sma05}.
\begin{figure*}[t]
 \includegraphics[width=12cm]{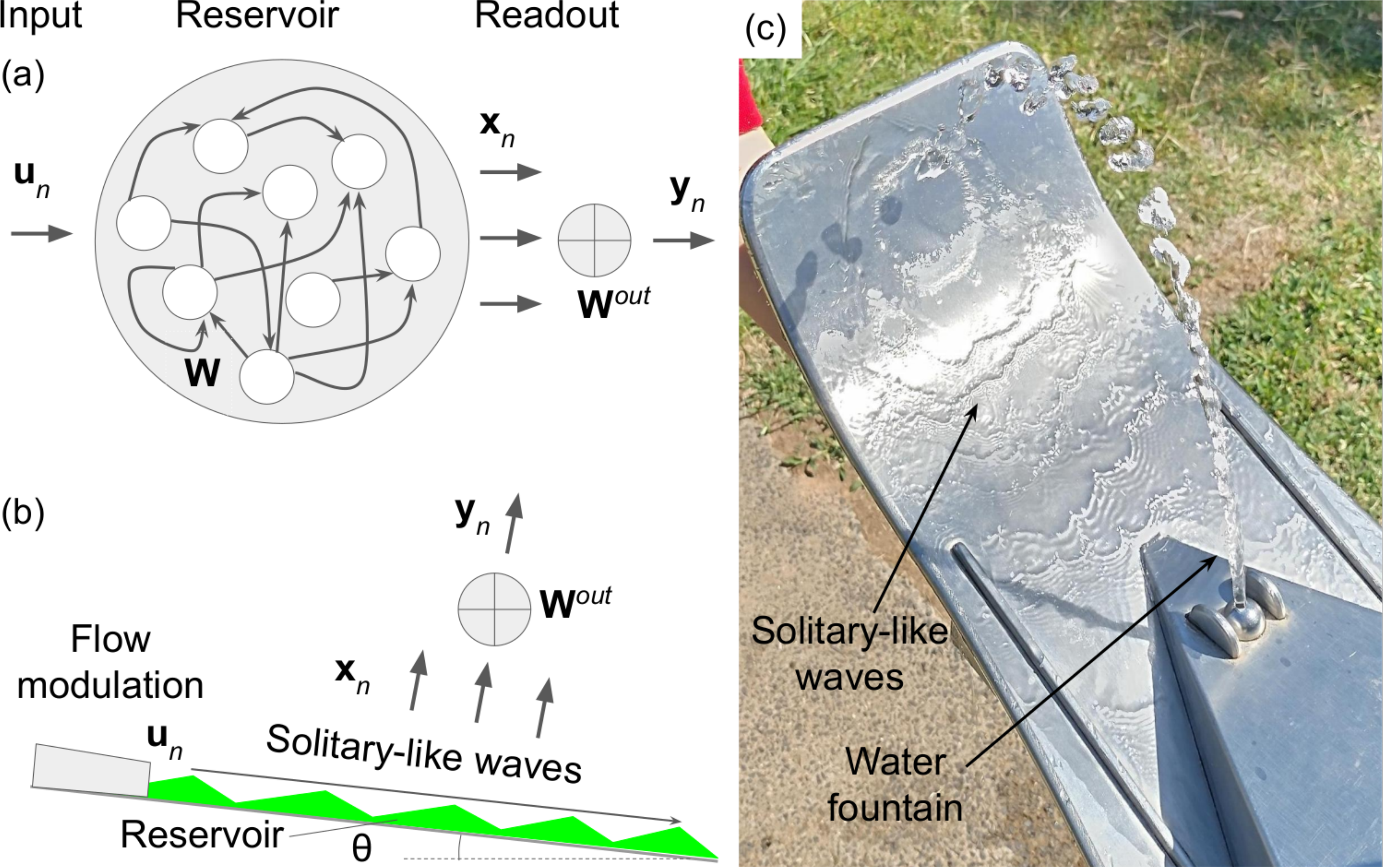}
 \caption{(a)~ESN-based RC algorithm.(b)~Sketch of the proposed SLRC. (c)~Photograph of SL waves of natural origin seen on the bowl of a drinking fountain that spouts water.\label{Fig1}}
\end{figure*}

The ability of an RC system to perform complex forecasting tasks originates from its structural organisation [Fig.~\ref{Fig1}(a)] that separates a fast linear readout layer from a much slower artificial neural network called the reservoir \cite{Luk09}. The reservoir consists of a network of randomly connected artificial neurons, and it converts a time-varying input into a spatio-temporal pattern of neural activations that is then transformed into practical outputs using the readout layer. While the readout layer is task-specific and memory-less, the universal reservoir is based on the principles of nonlinear dynamical systems and has certain memory \cite{Jae05}. Such a structure enables a reservoir consisting of a relatively small number of artificial neurons to be more efficient in resolving certain practical problems than a computer program run on a supercomputer. Thus, one can draw an analogy between a reservoir and some biological brains \cite{Maa02}, including the brains of fruit flies and mosquitos that, on average, have just about 200,000 neurons (in contrast to mammalian brains that have tens of billions neurons) but enable the insects to perform complicated tasks while they navigate and find food \cite{Raj21}.

Although in the framework of LSM the analogy between a reservoir and a liquid was made rather at an abstract level \cite{Maa02}, follow-up research works have demonstrated that physical liquids inherently possess all characteristics required for the creation of a reservoir \cite{Fer03, Jon07, Nak15, Nak20, Got21, Mak21_ESN, Mat22}. There have also been theoretical proposals \cite{Jun19, Zen20, Mar20, Sil21} of RC systems based on optical solitons \cite{Kiv03}. However, there have been no experimental demonstrations of soliton-based RC system yet.

In this paper, we experimentally demonstrate a physical RC system that exploits the dynamics of solitary-like (SL) waves [Fig.~\ref{Fig1}(b)] that originate from the spatio-temporal evolution of liquid films flowing over an inclined surface \cite{Liu94, Cha94, Kal12}. We confirm the plausibility of the SL waves based RC (SLRC) system using several standard test protocols designed to benchmark computer programs that implement the concept of RC. SL waves are ubiquitous in nature and can be readily observed in many everyday-life situations [Fig.~\ref{Fig1}(c)]. Subsequently, the principles of SLRC demonstrated in this work should be applicable to numerous physical system that support roll waves \cite{Bal04, Kal12}, including microfluidic devices \cite{Hu18}.

\section{Algorithmic reservoir computing\label{sec:2}}
A standard ESN algorithm [Fig.~\ref{Fig1}(a)] that we employ as a reference uses the following nonlinear update equation:
\begin{eqnarray}
  {\bf x}_{n} = (1-\alpha){\bf x}_{n-1}+\alpha\tanh({\bf W}^{in}{\bf u}_{n}+{\bf W}{\bf x}_{n-1})\,,
  \label{eq:RC1}
\end{eqnarray}
where $n$ is the index denoting equally-spaced discrete time instances $t_n$, ${\bf u}_n$ is the vector of $N_u$ input values, ${\bf x}_n$ is a vector of $N_x$ neural activations of the reservoir, the operator $\tanh(\cdot)$ is a sigmoid activation function of a neuron, ${\bf W}^{in}$ is the input matrix containing $N_x \times N_u$ elements, ${\bf W}$ is the recurrent weight matrix consisting of $N_x \times N_x$ elements and $\alpha \in (0, 1]$ is a parameter that controls the reservoir's temporal dynamics.

One calculates the output weights ${\bf W}^{out}$ by solving a system of linear equations ${\bf Y}^{target} = {\bf W}^{out}{\bf X}$, where ${\bf X}$ and ${\bf Y}^{target}$ are the state matrix and the target matrix that are constructed using, respectively, ${\bf x}_n$ and the vector of target outputs ${\bf y}_n^{target}$ as columns for each discrete time instant $t_n$. The solution is obtained in the form ${\bf W}^{out} = {\bf Y}^{target} {\bf X^\top} ({\bf X}{\bf X^\top} + \beta {\bf I})^{-1}$, where ${\bf I}$ is the identity matrix, $\beta \geq 0$ is a regularisation coefficient and ${\bf X^\top}$ is the transpose of ${\bf X}$ \cite{Luk09}. Then, one solves Eq.~(\ref{eq:RC1}) for new input data ${\bf u}_n$ and computes the output vector ${\bf y}_n={\bf W}^{out}[1;{\bf u}_n;{\bf x}_n]$ using a constant bias and the concatenation $[{\bf u}_n;{\bf x}_n]$.

\section{Solitary-like wave physical reservoir system}
Figure~\ref{Fig1}(b) shows a sketch of the experimental setup, where SL waves propagating over a metal plate inclined with respect to the ground by the angle $\theta=3$\,$^{o}$ \cite{Mak22} are used as a computational reservoir. The electronic modulation of the pump flow rate is used to control the thickness of the liquid film and to create SL waves on its surface. Adding an organic fluorescent dye to the liquid (tap water) and irradiating it with UV light, we generate green fluorescence light and measure variations of its intensity caused by the SL waves. The optical intensity profiles are recorded using an overhead high-speed (210\,frames per second) digital camera and then processed using customised software \cite{Mak22}.
\begin{figure*}[t]
 \includegraphics[width=18cm]{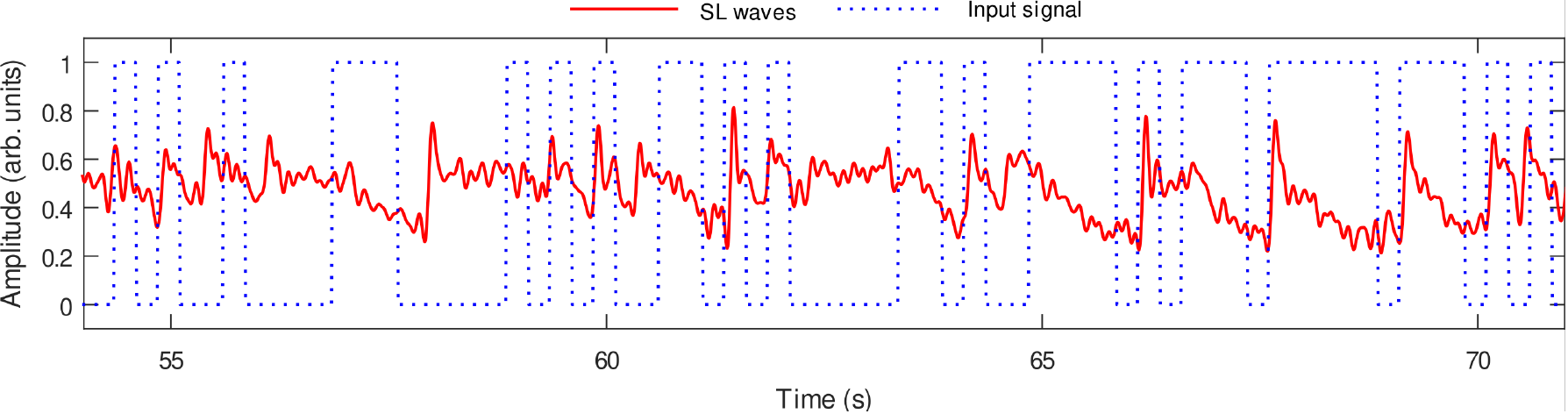}
 \caption{SL waves produced by a short training section of the randomly generated binary input signal, where each '1' and '0' value is represented as 0.25-second-long square pulses and 0.25-second-long gaps between them, respectively.\label{Fig2}}
\end{figure*}

To validate SLRC, we adopt the approach used to test the ability of ESN computer programs to predict chaotic time series (TS) \cite{Jae05, Gau21, Mak21_ESN}. First, the response of the SL waves to a sufficiently long data set is recorded in one single measurement. Then, the first half of the input data set is used for training of the RC system but its second half is reserved for tests. Thus, using the first half of the input data we calculate ${\bf W}^{out}$, but the second half the data set, ${\bf u}_n$, produces the neural activations ${\bf x}_n$ needed to compute the output as ${\bf y}_n={\bf W}^{out}[1;{\bf u}_n;{\bf x}_n]$.

As the first TS that is intended to test the memory capacity of SLRC, we use a randomly generated vector of binary '0' or '1' values. This data vector is converted into an electric signal using a digital function generator, where 0.25-second-long square pulses correspond to the '1' states but 0.25-second-long spaces between the pulses represent the '0' states. This electric signal modulates the pump flow [Fig.~\ref{Fig1}(b)]: the '1' states of the TS are implemented as a brief change in the flow rate that results in the generation of one SL wave but the '0' states correspond to no flow modulation. For example, an input data subset '10101' is represented by three 0.25-second-long pulses separated by two 0.25-second-long periods of no flow modulations, but '111001111' corresponds to one 0.75-second-long pulse separated by 0.5 seconds of no flow modulation from the following 1-second-long pulse. The chosen duration of both the pulses and the gaps between them is based on the optimal SL wave generation parameters established in \cite{Mak22}.

Figure~\ref{Fig2} shows the SL waves produced by the training portion of the random TS. While a periodic square pulse signal corresponding to a regular '101010...' binary sequence produces a pattern of equally spaced SL waves (not shown here, see \cite{Mak22} for a relevant discussion), the SL wave profile in Fig.~\ref{Fig2} follows the random pattern of the input TS. The SL waves do not precisely reproduce the training pulses, which is a typical picture seen at the training stage of physical RC systems \cite{Wat20, Mak21_ESN}. As also established in algorithmic RC systems \cite{Jae05}, the response of the reservoir must not be a copy of the training signal since in that case the trained RC system would not be able to produce an output that resembles the target TS.

The SL wave signal in Fig.~\ref{Fig2} corresponds to a single neural activation of the reservoir. To generate more neural activations, we sample this signal with a small discrete time step following the procedure described in \cite{Wat20}, thus producing 40 neural activations. We arrange the so-obtained activations into a state matrix ${\bf X}$ and use it to calculate ${\bf W}^{out}$ following the standard ESN algorithm [Fig.~\ref{Fig1}(a)].

As the second test, using the same training approach as above, we task the SLRC to predict a Mackey-Glass TS (MGTS). MGTS is often employed to test RC systems due to its nonlinear chaotic behaviour \cite{Thi03, Luk12, Mak21_ESN} and it is produced by the delay differential equation \cite{Mac77}
\begin{eqnarray}
  \dot{x}_{_{MG}}(t)
  &=&\beta_{_{MG}}\frac{x_{_{MG}}(\tau_{_{MG}}-t)}{1+x_{_{MG}}^{q}(\tau_{_{MG}}-t)}-\gamma_{_{MG}}x_{_{MG}}(t)\,,
  \label{eq:MG}
\end{eqnarray}
where we choose $\tau_{_{MG}}=17$ and set $q=10$, $\beta_{_{MG}}=0.2$ and $\gamma_{_{MG}}=0.1$. The resulting TS modulates liquid flow in the experiment. However, in contract to the square pulses, the MGTS modulation is continuous. We note that the chosen set of MGTS parameters is often used to test the performance of standard ML algorithms, including autoregressive moving average (ARMA) \cite{Thi03}. Algorithmic RC systems are known to outperform ARMA in forecasting chaotic TS \cite{Mac14}. Thus, we also use the MGTS test to verify whether SLRC could outperform ARMA.

\section{Benchmark test}
{\it{Memory capacity.}} A reservoir suitable for applications in RC should have an internal memory \cite{Jae05}. We test the memory capacity of SLRC using the $k$-delay task applied to the random binary TS, where we calculate the correlation coefficient $r({\bf y}^{target}_{n-k}, {\bf y}_{k,n})$ between a delayed target signal ${\bf y}^{target}_{n-k}$ and the predicted output ${\bf y}_{k,n}$ produced by the RC system trained using the training data ${\bf u}_{k,n}$ with a discrete time delay $k$ \cite{Jae05}. The function $r^{2}({\bf y}^{target}_{n-k}, {\bf y}_{k,n})$ can accept any value from 0 to 1, being 0 an indicator of the full loss of correlation and 1 corresponding a 100\% correlation between the correct target data and the prediction made by the RC system. Calculating $r^2$ as a function of the discrete delay time $k$, we determine the memory capacity (MC) as
\begin{eqnarray}
  MC = \sum_{k=1}^{k_{max}} r^{2}({\bf y}^{target}_{n-k}, {\bf y}_{k,n})\,,
  \label{eq:RC5}
\end{eqnarray}
being $k_{max}$ the maximum delay considered in the analysis.

\begin{figure}[t]
 \includegraphics[width=8.5cm]{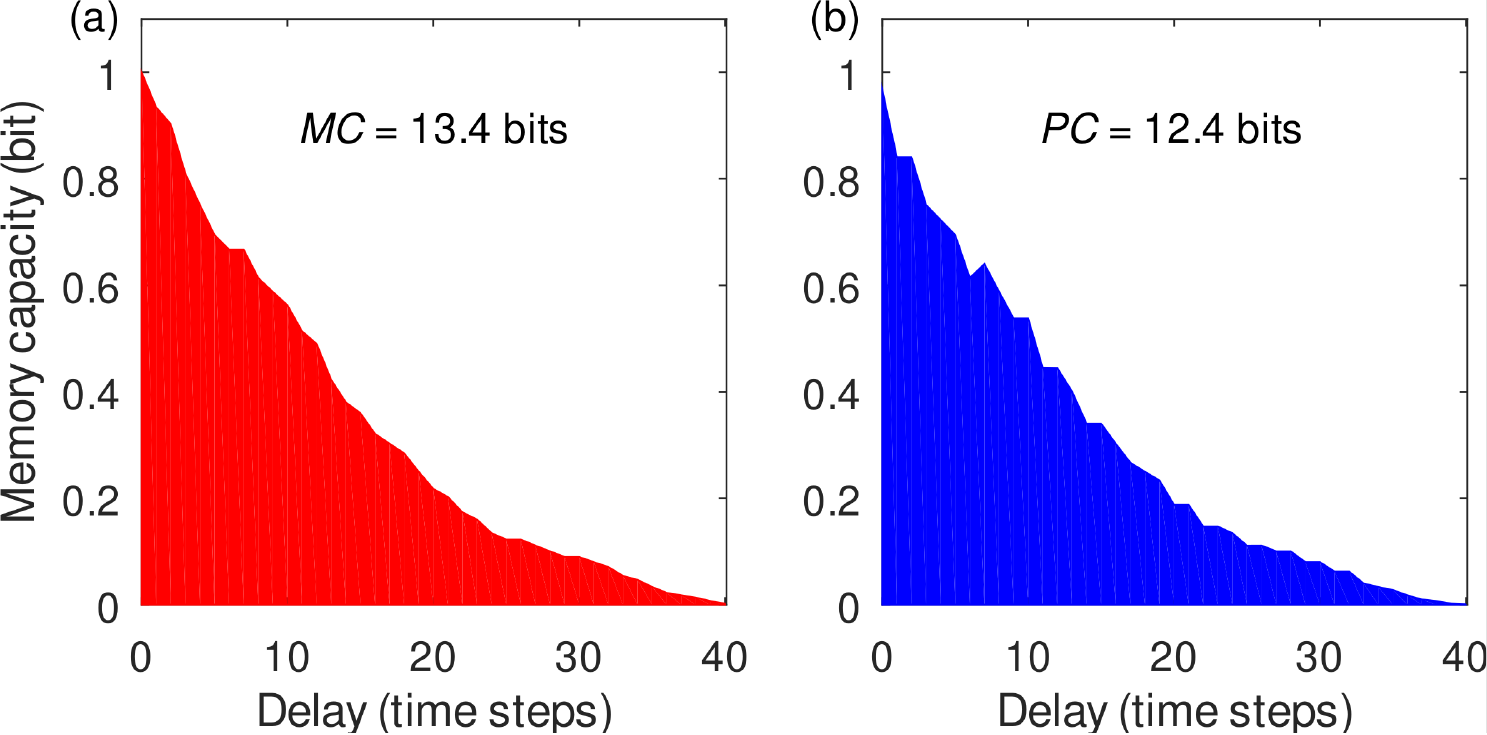}
 \caption{Memory capacity of the SLRC determined using (a)~$k$-delay and (b)~parity check tasks.\label{Fig3}}
\end{figure}

Figure~\ref{Fig3}(a) plots $r^{2}$ for $k_{max}=20$ and produces $MC = 13.4$\,bits. We observe that the last non-zero value of $r^2$ corresponds to $k=40$ thus showing that SLRC, which has $N=40$ neurons, retains some memory of the past inputs up to the theoretical fundamental limit $MC=N$ of algorithmic RC systems \cite{Jae05}. Yet, in practice, in algorithmic RC systems $MC \approx 0.3N$ due to an accumulated effect of rounding errors in the network update equations \cite{Jae05}. Obtaining $N/MC \approx 0.35$ for SLRC, we conclude that imperfections of the experimental setup and of the image processing technique used to detect SL waves play the role of rounding error in an algorithmic RC system.

The value of MC reported in Fig.~\ref{Fig3}(a) is an order of magnitude higher than that reported in \cite{Zen20}, where an optical soliton RC system was theoretically suggested. We hypothesise that the application of SL waves in the field of RC since may be advantageous since they possess unique nonlinear properties that are not available with solitary wave of other physical nature. Indeed, SL waves can merge \cite{Liu94} instead of passing one through another, with the latter being the case of well-known optical \cite{Kiv03} and matter \cite{Rem94} soliton waves, which we speculate may increase the memory of SLRC to the past inputs.

To support our arguments, we first refer to the literature on the memory capacity of RC systems and we confirm that the behaviour of $r^2$ observed in Fig.~\ref{Fig3}(a) closely resembles that of reservoirs consisting of artificial neurons with highly nonlinear activation functions \cite{Gun08, Ver19}. We also task the SLRC with a parity check test used to evaluate the performance of recurrent neural network that operate in a nearly chaotic dynamic regime \cite{Ber04}. In this test, the RC system predicts a TS defined as $PARITY({\bf u}_{n-k}, {\bf u}_{n-k-1}, {\bf u}_{n-k-2})$ for increasing delays $k$. The so-generated TS is a binary sequence itself and its forecast is a challenging task for any RC system since the parity function is not linearly separable and its prediction requires the reservoir to have memory. Similarly to Eq.~(\ref{eq:RC5}), we calculate PC memory capacity by summation of the parity check values up to the maximum delay $k_{max}$.

Figure~\ref{Fig3}(b) shows the outcome of the parity check test that produces $PC = 12.4$\,bits. In the literature on physical RC systems, the value of $PC$ is usually slightly smaller than the value of $MC$ and, typically, it reaches 2\,bits \cite{Fur18, Wat20, Lia21, Mat22}. The value of $PC$ for algorithmic RC systems is larger than that of physical RC systems \cite{Ber04}. While for the SLRC the value of $PC$ is also slightly smaller than that of $MC$, it is six times larger than the typical value for physical RC systems used in this work as a reference.

A plausible explanation of this difference may be the fact that the phase space of a standard algorithmic reservoir is chosen so that is does not contain multiple periodic or chaotic attractors or fixed points, the presence of which may disrupt the echo state property \cite{Luk12}. Physical RC systems are also often designed according to this rule. However, in practice, many physical RC systems with complex dynamics not only can reproduce the behaviour of algorithmic RC systems having an echo state property but they can also resolve complex nonlinear problems with a higher efficiency \cite{Ber04, Mak21_ESN}. Since SL waves exhibit a complex nonlinear behaviour \cite{Liu94}, similarly to the RC systems discussed in \cite{Ber04, Mak21_ESN} SLRC should be capable of resolving some complex nonlinear problems with high efficiency.

We also note that in Fig.~\ref{Fig3}(b) at the zero delay the parity check produces the memory of approximately $0.96$\,bit instead of the theoretically expected 1\,bit corresponding to 100\% correlation. The exact origin of this discrepancy in our proof-of-principle experiment is unknown. However, the following explanations may be plausible. Firstly, it may be an artefact since it is not observed in the MC test. Secondly, in algorithmic RC systems, this situation is known to originate from long-term reverberations in the reservoir that deteriorate the recall of immediately past inputs \cite{Kir91, Jae05}. In the SLRC, reverberations may be caused by technical imperfections of the inlet fixture and by small reflections of SL from the downstream edge of the inclined plate, which both are well-established facts \cite{Mak22}.
\begin{figure*}[t]
 \includegraphics[width=18cm]{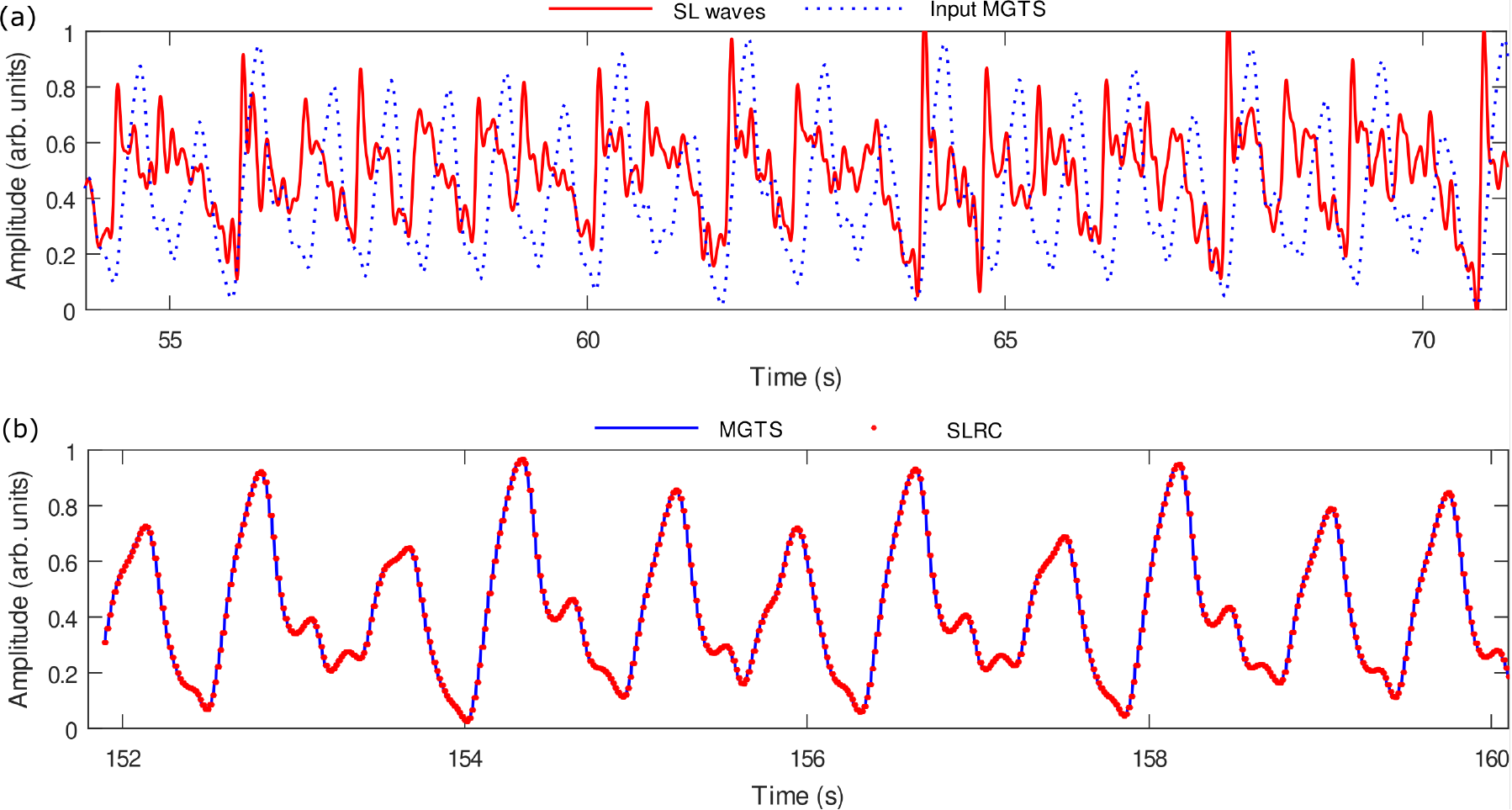}
 \caption{MGTS test of SLRC. (a)~SL waves produced by a short training section of the MGTS input signal. (b)~One-timestep ahead prediction made by SLRC (dots) compared with the expected target signal (line). Note the different timescales use to plot the training and prediction stage data.\label{Fig4}}
\end{figure*}

{\it{Mackey-Glass time series}.}
We test the ability of SLRC to predict MGTS using a one-step ahead prediction tasks that is often employed to test the accuracy of neural network models \cite{Mar04, Rod10, Mor21}. In this task, the value of a TS at a time step $t_n$ is mapped to a future time step $t_{n+1}$ and the resulting prediction is compared with the expected target signal to calculate the accuracy of the reservoir.

Using the approach outlined in \cite{Mak21_ESN}, we resample the MGTS signal so that the frequency of the principal peak in its Fourier spectrum equals 2\,Hz, which is the frequency of SL waves investigated in the relevant experimental work \cite{Mak22}. In Fig.~\ref{Fig4}(a) we plot a section of the training MGTS signal and the traces of the SL waves created by it. As with the square pulses in Fig.~\ref{Fig2}, we observe an SL wave pattern that follows but not precisely fit the training MGTS, which indicates that the trained RC system will be able to generalise the learnt input signal patterns \cite{Jae05}.

In Fig.~\ref{Fig4}(b), we can see that the SLRC correctly predicts the target MGTS. We calculate the normalised mean square error (NMSE), which is a standard measure of the accuracy of algorithmic RC systems \cite{Luk09}, as
\begin{eqnarray}
  NMSE({\bf y}, {\bf y}^{target}) = \frac{\Bigl \langle\left\Vert {\bf y}_{n}-{\bf y}_{n}^{target} \right\Vert^{2} \Bigl \rangle}{\Bigl \langle\left\Vert {\bf y}_{n}-\langle{\bf y}_{n}^{target}\rangle \right\Vert^{2} \Bigl \rangle}\,,
  \label{eq:RC4}
\end{eqnarray}
where $\left\Vert \cdot \right\Vert$ is the norm and $\langle \cdot \rangle$ denotes the averaging over time. An $NMSE=0$ means that the RC system produces a 100\% correct prediction but $NMSE=1$ means that the RC system has fully failed to predict the target data. We obtain $NMSE_{SLRC} \approx 6.6 \times 10^{-4}$. Then, we task software that implements the ESN model Eq.~(\ref{eq:RC1}) \cite{Luk12} to forecast MGTS and obtain $NMSE_{ESN} \approx 1.75 \times 10^{-7}$. A three orders of magnitude difference between NMSE of SLRC and ESN is inconsequential for our analysis and is attributed to a noise present in the SLRC measurements. We also use software that implements an ARMA model \cite{Kha23} and predicts MGTS in a one-step-ahead regime. We obtain $NMSE_{ARMA} \approx 1.5 \times 10^{-3}$, which shows that the accuracy of ARMA is lower than that of RC systems, which is an established fact \cite{Mac14}.

\section{Theoretical analysis}
It has been shown that considerable memory capacity of physical RC systems can be achieved under a condition, when the system's response to one pulse of a chaotic TS does not fully decay before it is presented with the following pulses, which enables the system to retain some traces of its response to several consecutive pulses of the input TS \cite{Wat20}. SLRC has a similar memory mechanism that is based on the property of SL waves of higher amplitude to overtake and absorb SL waves of lower amplitude \cite{Liu94}. We theoretically investigate this mechanism using a model, where, for the sake of illustration, we excite SL waves using three square pulses.

We employ a simplified hydrodynamic Shkadov model that is based on the boundary layer approximation of the Navier-Stokes equation under the assumption of the parabolic longitudinal velocity profile \cite{Shkadov67,Shkadov68} and that is formulated in terms of two coupled equations for the local film thickness $h(x,t)$ and the fluid flux across the layer $q(x,t)$
\begin{eqnarray}
\label{theq1}
\rho \left[ \partial_t q  +\frac{6}{5}\partial_x \left(\frac{q^2}{h}\right)\right]&=&-\frac{3\mu q}{h^2}+\sigma h\partial_x^3 h\nonumber\\
&-&\rho g \cos(\theta)h\partial_x h+\rho g\sin(\theta)h,\nonumber\\
\partial_t h +\partial_x q&=&f(x,t),
\end{eqnarray}
where $\theta$ is the inclination angle, $\mu$ the dynamic viscosity and $\sigma$ the liquid-air surface tension. The term $f(x,t)$ in the second equation in Eqs.\,(\ref{theq1}) is introduced to mimic the time-dependent influx of the fluid induced by the inlet. We use a Gaussian $f(x,t)=\exp{(-a(x-x_0)^2)}s(t)$ centred around the location of the inlet $x_0$ with parameter $a$ chosen sufficiently large to ensure that the width of the Gaussian is significantly smaller than the typical length of the surface waves. The binary function $s(t)=f_a\{0,1\}$ is introduced to describe the time modulation of the fluid influx and $f_a$ stands for the amplitude of the influx. We numerically integrate Eqs.\,(\ref{theq1}) in the $(-L\leq x\leq L)$ domain of total length of $2L=400$\,cm using periodic boundaries. The inlet is located at $x_0=100$\,cm. The relatively large domain length enables us to observe the dynamics of the waves over a time period of up to 15\,seconds, before the waves travel over the distance of $2L$.

\begin{figure*}[t]
 \includegraphics[width=18cm]{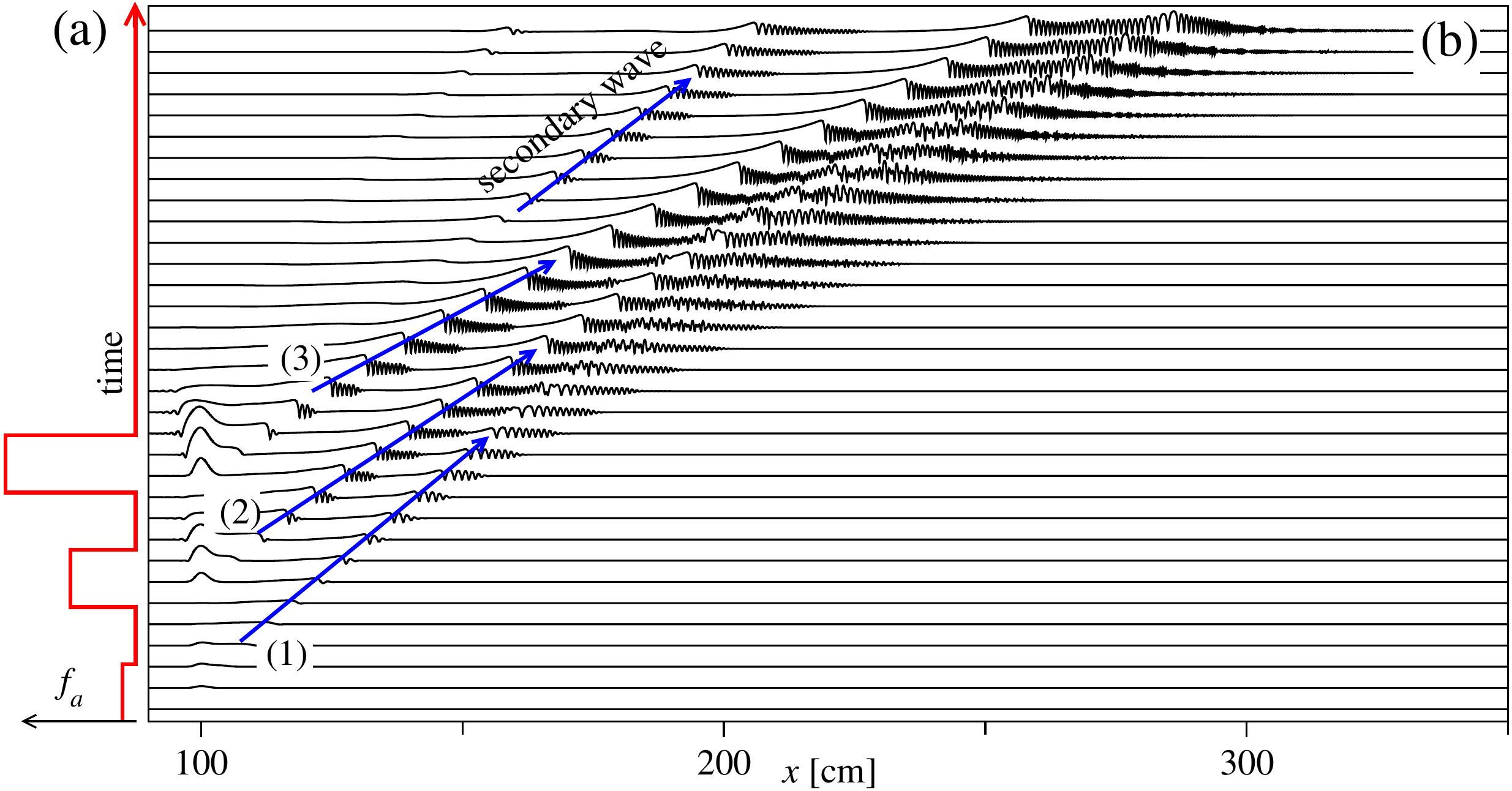}
 \caption{(a)~Modulation function $s(t)$ used to excite three SL waves with different amplitudes. (b)~Space-time plot of the temporal evolution of the three SL waves, labelled by the arrows $(1,2,3)$, excited by the modulation function shown in panel~(a). The first smaller wave $(1)$ propagates slower than the second and the third larger waves $(2,3)$. The second wave inelastically collides with the first wave about 100\,cm further downstream.\label{Fig5}}
\end{figure*}

To demonstrate the interaction and collision of waves with different amplitudes, we excite three SL waves on a 0.7-mm-thick water film using three flow modulation phases with a gradually increasing amplitude. All three phases have a duration of 0.5\,seconds and they are separated one from another by 0.5\,seconds of no flow modulation [Fig.~\ref{Fig5}(a)]. The space-time plot of the excited waves is shown in Fig.~\ref{Fig5}(b), where the arrows labelled as $(1,2,3)$ highlight the propagation direction of the three SL waves. As anticipated, the first smaller wave propagates slower than the second larger wave. These two waves eventually collide approximately 100\,cm further downstream and form a new wave with a complex profile. Such an interaction scenario is known as an inelastic wave collision \cite{Liu94}. All three waves $(1,2,3)$ eventually form one compact structure that consists of many smaller waves around 150\,cm further downstream from the inlet location.

Thus, we can see that, when the SL waves are detected very close to the inlet, they do not have enough time to develop and interact, which, in turn, means that the memory capacity of SLRC is low. The memory capacity will be low also when SL waves are detected far downstream, where a significant transformation of the waveform implies that the ability of the reservoir to memorise the input signal is compromised. Subsequently, it is plausible that the optimal memory capacity will be achieved when SL waves are detected midway the inlet and the far downstream region where the SL waves merge, which is the approach adopted in the experiments conducted in this work.

\section{Discussion}
Before we conclude, we discuss several potential improvements of SLRC that can be done in the future. Firstly, the experimental setup used in this work holds the potential to be extended to enable measurements, where the RC system could generate a new TS using its own predictions as the input data. Such an operating regime is called the generative mode and its implementation is a challenging task for both physical and algorithmic RC systems \cite{Jae05, Luk12, Mak21_ESN}. In particular, the operation in a generative mode requires introducing a feedback circuit that converts the output of the readout layer into new input data at every discrete time step. However, transfer of data from the readout layer, which, typically, is a computer program itself, to the input layer of the RC system introduces a time delay that can be several orders of magnitude larger than the timescale of the reservoir dynamics. For instance, if the reservoir's natural frequency is of order of several GHz \cite{Wat20}, all readout calculations must be done at the GHz scale or above. In the case of SLRC, where the frequencies of SL waves are of order of several Hz, this problem can be resolved using a laser vibrometry technique to detect the SL waves \cite{Mak19_Faraday} and then processing the optical signal using ultrafast opto-electronic and photonics systems that have already been adopted in the field of ML \cite{Sor20, Che20_review}. The time delay introduced by such a processing scheme will be negligibly small compared with the temporal dynamics of SL waves.

Apart from the operation in the generative mode, the introduction of a time delay between the output and input of the reservoir may help increase the memory capacity of the RC system \cite{Rio19, Wat20, Kon22}, also enabling the reservoir to operate using just a single artificial neuron \cite{App11}. Yet, in the field of fluid dynamics, the introduction of a feedback control has been used to stabilise liquid film flows \cite{Tho16}, which may also help improve the performance of SLRC.

\section{Conclusions\label{sec:conclusions}}
We have experimentally demonstrated a physical computational reservoir system enabled by solitary-like waves propagating on the surface of a liquid thin film flowing over an inclined surface. While computational reservoirs based on optical and matter-wave solitons have been theoretically predicted but not experimentally demonstrated yet, solitary wave dynamics of liquid films has unique features that we have employed in the field of machine learning for the first time. We trained and exploited the solitary-like wave reservoir to perform several essential benchmark tests used to evaluate the performance of machine learning algorithms designed for digital computer, showing that the solitary-like wave reservoir has considerable memory capacity and can predict chaotic time series more efficiently than certain standard algorithmic machine learning models.


\bibliographystyle{eplbib}
\bibliography{epl_maksymov_pototsky_v1}

\begin{thebibliography}{10}
\expandafter\ifx\csname url\endcsname\relax\def\url#1{\texttt{#1}}\fi

\bibitem{Nak21}
\Name{Nakajima K. \and Fisher I.} \Book{Reservoir Computing} (Springer, Berlin)
  2021.

\bibitem{Kir91}
\Name{Kirby K.~G.} \REVIEW{Proc.~Florida AI Research Symposium
  (FLAIRS)}{}{1991}{66}.

\bibitem{Maa02}
\Name{Maass W., Natschl{\"a}ger T. \and Markram H.} \REVIEW{Neural
  Comput.}{14}{2002}{2531}.

\bibitem{Luk09}
\Name{Luko{\v s}evi{\v c}ius M. \and Jaeger H.}
  \REVIEW{Comput.~Sci.~Rev.}{3}{2009}{127}.

\bibitem{Gau21}
\Name{Gauthier D.~J., Bollt E., Griffith A. \and Barbosa W.~A.~S.}
  \REVIEW{Nat.~Commun.}{12}{2021}{5564}.

\bibitem{Mar04}
\Name{Marcellino M., Stock J.~H. \and Watson M.~W.}
  \REVIEW{J.~Econom.}{135}{2006}{499}.

\bibitem{Sma05}
\Name{Small M.} \Book{Applied Nonlinear Time Series Analysis: Applications in
  Physics, Physiology and Finance} (World Scientific, Singapore) 2005.

\bibitem{Jae05}
\Name{Jaeger H.} \Book{A tutorial on training recurrent neural networks,
  covering {BPPT, RTRL, EKF} and the “echo state network” approach} ({GMD}
  Report 159, German National Research Center for Information Technology) 2005.

\bibitem{Raj21}
\Name{Raji J.~I. \and Potter C.~J.} \REVIEW{PLOS ONE}{16}{2021}{1}.

\bibitem{Fer03}
\Name{Fernando C. \and Sojakka S.} \Book{Pattern recognition in a bucket} in
  proc. of \Book{Advances in Artificial Life}, edited by \Name{Banzhaf W.,
  Ziegler J., Christaller T., Dittrich P. \and Kim J.~T.} (Springer, Berlin)
  2003 pp. 588--597.

\bibitem{Jon07}
\Name{Jones B., Stekel D., Rowe J. \and Fernando C.} \Book{Is there a liquid
  state machine in the bacterium escherichia coli?} in proc. of \Book{2007 IEEE
  Symposium on Artificial Life} 2007 pp. 187--191.

\bibitem{Nak15}
\Name{Nakajima K. \and Aoyagi T.} \REVIEW{IEICE Tech.~Rep.}{115}{2015}{109}.

\bibitem{Nak20}
\Name{Nakajima K.} \REVIEW{Jpn.~J.~Appl.~Phys.}{59}{2020}{060501}.

\bibitem{Got21}
\Name{Goto K., Nakajima K. \and Notsu H.}
  \REVIEW{N.~J.~Phys.}{23}{2021}{063051}.

\bibitem{Mak21_ESN}
\Name{Maksymov I.~S., Pototsky A. \and Suslov S.~A.}
  \REVIEW{Phys.~Rev.~E}{105}{2021}{044206}.

\bibitem{Mat22}
\Name{Matsuo T., Sato D., Koh S.-G., Shima H., Naitoh Y., Akinaga H., Itoh T.,
  Nokami T., Kobayashi M. \and Kinoshita K.} \REVIEW{ACS
  Appl.~Mater.~Interfaces}{14}{2022}{36890}.

\bibitem{Jun19}
\Name{Jiang J. \and Lai Y.-C.} \REVIEW{Phys.~Rev.~Research}{1}{2019}{033056}.

\bibitem{Zen20}
\Name{Zeng Q., Wu Z., Yue D., Tan X., Tao J. \and Xia G.}
  \REVIEW{Appl.~Opt.}{59}{2020}{6932}.

\bibitem{Mar20}
\Name{Marcucci G., Pierangeli D. \and Conti C.}
  \REVIEW{Phys.~Rev.~Lett.}{125}{2020}{093901}.

\bibitem{Sil21}
\Name{Silva N.~A., Ferreira T.~D. \and Guerreiro A.} \REVIEW{New
  J.~Phys.}{23}{2021}{023013}.

\bibitem{Kiv03}
\Name{Kivshar Y.~S. \and Agrawal G.~P.} \Book{Optical Solitons:~From Fibers to
  Photonic Crystals} (Academic Press, New York) 2003.

\bibitem{Liu94}
\Name{Liu J. \and Gollub J.~P.} \REVIEW{Phys.~Fluids}{6}{1994}{1702}.

\bibitem{Cha94}
\Name{Chang H.-C.} \REVIEW{Annu.~Rev.~Fluid Mech.}{26}{1994}{103}.

\bibitem{Kal12}
\Name{Kalliadasis S., Ruyer-Quil C., Scheid B. \and Velarde M.~G.}
  \Book{Falling Liquid Films} (Springer-Verlag, London) 2012.

\bibitem{Bal04}
\Name{Balmforth N.~J. \and Mandre S.} \REVIEW{J.~Fluid Mech.}{514}{2004}{1}.

\bibitem{Hu18}
\Name{Hu X. \and Cubaud T.} \REVIEW{Phys.~Rev.~Lett.}{21}{2018}{044502}.

\bibitem{Mak22}
\Name{Maksymov I.~S. \and Pototsky A.} \REVIEW{Appl.~Sci.}{13}{2022}{1888}.

\bibitem{Wat20}
\Name{Watt S. \and Kostylev M.} \REVIEW{Phys.~Rev.~Appl.}{13}{2020}{034057}.

\bibitem{Thi03}
\Name{Thissen U., {van Brakel} R., {de Weijer} A.~P., Melssen W.~J. \and
  Buydens L.~M.~C.} \REVIEW{Chemom.~Intell.~Lab.~Syst.}{69}{2003}{35}.

\bibitem{Luk12}
\Name{Luko{\v s}evi{\v c}ius M.} \Book{A practical guide to applying echo state
  networks} in \Book{Neural Networks: Tricks of the Trade, Reloaded}, edited by
  \Name{Montavon G., Orr G.~B. \and M{\" u}ller K.-R.} (Springer, Berlin) 2012
  pp. 659--686.

\bibitem{Mac77}
\Name{Mackey M.~C. \and Glass L.} \REVIEW{Science}{197}{1977}{287}.

\bibitem{Mac14}
\Name{Maciel L., Gomide F., Santos D. \and Ballini R.} \Book{Exchange rate
  forecasting using echo state networks for trading strategies} in proc. of
  \Book{2014 IEEE Conference on Computational Intelligence for Financial
  Engineering and Economics (CIFEr)} 2014 pp. 40--47.

\bibitem{Rem94}
\Name{Remoissenet M.} \Book{Waves Called Solitons: Concepts and Experiments}
  (Springer) 1994.

\bibitem{Gun08}
\Name{Ganguli S., Huh D. \and Sompolinsky H.} \REVIEW{PNAS}{105}{2008}{18970}.

\bibitem{Ver19}
\Name{Verzelli P., Alippi C. \and Livi L.} \REVIEW{Sci.~Reps.}{9}{2019}{13887}.

\bibitem{Ber04}
\Name{Bertschinger N. \and Natschl{\"a}ger T.} \REVIEW{Neural
  Comput.}{16}{2004}{1413}.

\bibitem{Fur18}
\Name{Furuta T., Fujii K., Nakajima K., Tsunegi S., Kubota H., Suzuki Y. \and
  Miwa S.} \REVIEW{Phys.~Rev.~Appl.}{10}{2018}{034063}.

\bibitem{Lia21}
\Name{Liao Z., Wang Z., Yamahara H. \and Tabata H.} \REVIEW{Chaos Solitons
  Fract.}{153}{2021}{111503}.

\bibitem{Rod10}
\Name{Rodan A. \and Tino P.} \REVIEW{IEEE Trans.~Neural Netw.}{22}{2011}{131}.

\bibitem{Mor21}
\Name{Morales G.~B., Mirasso C.~R. \and Soriano M.~C.}
  \REVIEW{Neurocomputing}{461}{2021}{705}.

\bibitem{Kha23}
\Name{Khan S.} \Book{Mackey Glass Time Series Prediction Using Least Mean
  Square} (Matlab Central File Exchange) 2023.

\bibitem{Shkadov67}
\Name{Shkadov V.~Y.} \REVIEW{Izv.~Akad.~Nauk SSSR,
  Mekh.~Zhidk.~Gaza}{1}{1967}{43}.

\bibitem{Shkadov68}
\Name{Shkadov V.~Y.} \REVIEW{Izv.~Akad.~Nauk SSSR,
  Mekh.~Zhidk.~Gaza}{2}{1968}{20}.

\bibitem{Mak19_Faraday}
\Name{Maksymov I. \and Pototsky A.} \REVIEW{Phys.~Rev.~E}{100}{2019}{}.

\bibitem{Sor20}
\Name{Sorokina M.} \REVIEW{J.~Phys.~Photonics}{2}{2020}{044006}.

\bibitem{Che20_review}
\Name{Chembo Y.~K.} \REVIEW{Chaos}{30}{2020}{013111}.

\bibitem{Rio19}
\Name{Riou M., Torrejon J., Garitaine B., Araujo F.~A., Bortolotti P., Cros V.,
  Tsunegi S., Yakushiji K., Fukushima A., Kubota H., Yuasa S., Querlioz D.,
  Stiles M.~D. \and Grollier J.} \REVIEW{Phys.~Rep.~Appl.}{12}{2019}{024049}.

\bibitem{Kon22}
\Name{Kondrashov A.~V., Nikitin A.~A., Nikitin A.~A., Kostylev M. \and Ustinov
  A.~B.} \REVIEW{J.~Magn.~Magn.~Mater.}{563}{2022}{169968}.

\bibitem{App11}
\Name{Appeltant L., Soriano M.~C., der Sande G.~V., Danckaert J., Massar S.,
  Dambre J., Schrauwen B., Mirasso C.~R. \and Fischer I.}
  \REVIEW{Nat.~Commun.}{2}{2011}{468}.

\bibitem{Tho16}
\Name{Thompson A.~B., Gomes S.~N., Pavliotis G.~A. \and Papageorgiou D.~T.}
  \REVIEW{Phys.~Fluids}{28}{2016}{012107}.

\end{thebibliography}

\end{document}